# Appearance of Superconductivity in layered LaO$_{0.5}$F$_{0.5}$Bi$_2$

V.P.S. Awana[a*], Anuj Kumar[a], Rajveer Jha[a], Shiva Kumar Singh[a], Anand Pal[a], Shruti[b], J. Saha[b], S. Patnaik[b]

[a]Quantum Phenomena and Applications Division, National Physical Laboratory (CSIR), Dr. K. S. Krishnan Road, New Delhi-110012, India

[b]School of Physical Sciences, Jawaharlal Nehru University, New Delhi-110067, India

**Abstract**

Phase pure samples of LaOBiS$_2$ and LaO$_{0.5}$F$_{0.5}$BiS$_2$ are synthesized by conventional solid state reaction route via vacuum encapsulation technique at 800$^o$C for 12h. Both the samples are crystallized in tetragonal *P4/nmm* space group with lattice parameters $a$ = 4.066(1) Å, $c$ = 13.862(4) Å for LaOBiS$_2$; $a$ = 4.069(5) Å, $c$ = 13.366(2) Å for LaO$_{0.5}$F$_{0.5}$BiS$_2$. Bulk superconductivity is confirmed in LaO$_{0.5}$F$_{0.5}$BiS$_2$ with superconducting transition temperature ($T_c$) of 2.7K by *DC* magnetization and resistivity measurements. The Isothermal magnetization (*M-H*) measurement shows closed loops with clear signatures of flux pinning and irreversible behavior. The magneto-transport $\rho(T,H)$ measurements show resistive broadening and decrease in $T_c$ ($\rho$=0) to the lower temperature with increasing magnetic field. The magnetic phase diagram involving upper critical and irreversibility fields as a function of temperature has been ascertained. The upper critical field $H_{c2}$(0) is estimated to be ~19kOe corresponding to a Ginzburg-Landau coherence length of ~130 Oe.

**Corresponding Author**
* Dr. V. P. S. Awana, Senior Scientist
Quantum Phenomena and Applications Division
National Physical Laboratory (CSIR)
Dr. K. S. Krishnan Road, New Delhi-110012, India
E-mail: awana@mail.npindia.org
Ph. +91-11-45609357, Fax-+91-11-45609310

1. **Introduction**

The apparent commonalities among cuprates, pnictides, and heavy fermion superconductors have reorganized the material guidelines in the search of new high $T_c$ superconducting systems. Very recent example of the same is $LaO_{0.5}F_{0.5}BiS_2$ [1]. In this respect, significant attention has been paid towards ZrCuSiAs type compounds, particularly since the discovery of superconductivity in F-doped LaFeAsO with $T_c$ ~26 K [2]. Maintaining the basic structural unit of FeAs, couple of new systems have also been reported recently, such as $CeNi_{0.8}Bi_2$ [3] and doped $LaCo_2B_2$ [4] with superconducting transition temperature at around 4K. These compounds mimic the superconducting characteristics of layered $CuO_2$ and FeAs based High $T_c$ superconductors. Moreover, the very recent observation of superconductivity around 5.5 K in $NdO_{1-x}F_xBiS_2$ ($x = 0.10$-$0.70$) [5, 6] has once again emphasized the possibility of a new series of superconductors based on layered sulfides. In fact, this has already generated a great deal of interest in material science community [7, 8]. Various rare earth substitutions are being done to induce/enhance $T_c$ [7, 8]. The scaling of $T_c$ can be seen with rare earth ionic size variation [1, 5-8] *i.e.* it is increasing with decrease in rare earth ionic size. Further, the doping mechanism in $BiS_2$ based superconductors appears to be similar to that of cuprates and pnictides [9]. In this article, we report the synthesis and basic physical property characterization of $LaO_{1-x}F_xBiS_2$ ($x = 0.0$ and $0.50$) compounds. Both the compositions crystallized in ZrCuSiAs type tetragonal structure with space group $P4/nmm$. While $x = 0.5$ composition exhibits bulk superconductivity with a transition temperature of 2.7K, the $x = 0.0$ composition has the insulating nature with a characteristic resistivity upturn. The superconductivity of $LaO_{0.5}F_{0.5}BiS_2$ compound is confirmed from both magnetization and transport measurements.

## 2. Experimental

The samples were synthesized through standard solid state reaction route via vacuum encapsulation. High purity La, Bi, S, LaF$_3$, and La$_2$O$_3$ were weighed in stoichiometric ratio and ground thoroughly in a glove box under high purity argon atmosphere. The powders were subsequently pelletized and vacuum-sealed ($10^{-3}$ Torr) in separate quartz tubes. Sealed quartz ampoules were placed in tube furnace and heat treated at 800$^0$C for 12h with the typical heating rate of 2$^o$C/min., and subsequently cooled down slowly over a span of 6h to the room temperature. X-ray diffraction (*XRD*) was performed at room temperature in the scattering angular (*2θ*) range from 10$^o$-80$^o$ in the steps of 0.02$^o$, using *Rigaku Diffractometer* with *Cu K$_α$* (*λ* = 1.54Å) radiation. Rietveld analysis was performed using the *FullProf* program. Detailed *DC* magnetization and transport measurements were performed on Physical Property Measurements System (*PPMS*-14T, *Cryogenics*).

## 3. Results and discussion

The synthesized samples are black in color and hard in nature. The Rietveld refined room temperature *XRD* pattern of LaOBiS$_2$ and LaO$_{0.5}$F$_{0.5}$BiS$_2$ samples are shown in Figure 1 (a). Rietveld refinement is carried out using ZrCuSiAs structure and Wyckoff positions. The samples are crystallized in tetragonal structure with space group *P4/nmm*. Rietveld refinement shows both the samples are free from any Bi impurity phase. Worth mentioning is the fact that some Bi and Bi$_2$S$_3$ impurity were seen in studied samples of ref.1. Our samples are nearly phase pure at least with in XRD detection limit and superconducting for x = 0.5. Refined structural parameters (lattice parameters, atomic coordinates and site occupancy), are shown in the Table 1.

The representative unit cell of the compound in *P4/nmm* space group crystallization is shown in Figure 1(b). The layered structure includes La$_2$O$_2$ (rare earth oxide layer) and BiS$_2$ (fluorite type) layers. Various atoms with their respective positions are indicated in Figure 1(b). Bismuth (Bi), Lanthanum (La), and Sulfur (S1 and S2) atoms occupy the *2c* (0.25, 0.25, *z*) site. On the other hand O/F atoms are at *2a* (0.75, 0.25, 0) site.

Interestingly, the carrier doping mechanism is same as that of reported LaFeAsO [2]. In LaFeAsO/F the carriers are doped from LaO/F to superconducting FeAs layer. Similarly, in $LaO_{0.5}F_{0.5}BiS_2$, mobile carriers are doped from LaO/F redox layer to superconducting $BiS_2$ layer. Considerable decrease in *c*-parameter (~ 0.5 Å) is observed with F doping at O site. Change in lattice parameter confirms the substitution by F at O site. Also the *004* peak shifts towards higher angle side in F doped sample.

The magnetization data as a function of temperature under applied magnetic field of 30Oe is shown in Figure 2. The magnetization measurements are performed in both *FC* (Field cooled) and *ZFC* (Zero-field-cooled) protocols. The $LaO_{0.5}F_{0.5}BiS_2$ compound shows sharp superconducting onset at 2.7K. The bifurcation of *FC* and *ZFC* below $T_c$ marks the irreversible region. The superconducting shielding fraction as evidenced from *ZFC* diamagnetic volume susceptibility is around 6%, which is comparable to that (~10%) as reported in ref. 1. In fact, the increase of volume fraction of superconductivity in these systems is yet warranted. The inset in Figure 2 shows isothermal *M-H* curve of the sample at 1.8K. The *MH* plot is indicative of type-II behavior and mixed state is clearly seen. Summarily, the *FC/ZFC* magnetization and isothermal magnetization *M-H* data confirm bulk superconductivity in studied $LaO_{0.5}F_{0.5}BiS_2$.

Figure 3 shows the resistivity versus temperature (*ρ-T*) measurements in zero field and in presence of external magnetic field. Cooling from room temperature, the resistance of the sample increases with decreasing temperature and undergoes superconducting transition with an onset $T_c$ ~2.7K and $T_c$ (*ρ*=0) at 2.4K. The zero field resistivity of parent and F doped samples up to room temperature is shown in the upper inset. The pristine compound exhibits a resistivity upturn similar to what is seen in oxy-pnictides the origin of which is under investigation.

On the other hand normal state conduction of doped compound is evidently like a semiconductor that implies that the electron doping is not optimum and probably higher $T_c$ is achievable. With the application of magnetic field both the onset and offset $T_c$ decrease to lower temperature. Defining the intersection point between normal state resistivity extrapolation and superconducting transition line as the upper critical field $H_{c2}(T)$ and the intersection of superconducting line with the temperature axis as the irreversibility field $H^*$, we have estimated both $H_{c2}(T)$ and $H_{irr}(T)$ as a function of temperature. This is shown in the lower inset of Figure 3.

We note that $H_{c2}(0)$ represents complete destruction of superconductivity where as $H^*$ denotes onset of resistivity in the mixed state. The slope $dH_{c2}/dT|_{Tc}$ is estimated to be ~ 10kOe/K that implies *WHH* (Werthamer-Helfand-Hohenberg) $H_{c2}(0)$ (= - 0.69 $T_c$ $dH_{c2}/dT|_{Tc}$) value of 19kOe. From this, the Ginzburg-Landau coherence length ($\xi = 2.07*10^{-7}/2\pi H_{c2})^{1/2}$ is estimated to be ~ 130Å. Unlike the other new $BiS_2$ superconductor i.e., $Bi_4O_4S_3$ [10-11], positive magneto-resistance is not observed in normal state of $LaO_{0.5}F_{0.5}BiS_2$. This is because the reported $Bi_4O_4S_3$ [10-11] samples were contaminated with Bi impurity.

In Figure 4 we plot the Hall resistivity as a function of magnetic field at $T$ = 10K. Unlike $Bi_4O_4S_3$, a linear response is indicated in this relatively low field regime. The conduction mechanism is dominated by electrons and the calculated Hall coefficient and carrier concentration at 10 K are $n$ = 1.24 x$10^{20}$ per $cm^3$ and $R_H$ = 5.04036 x $10^{-8}$ $m^3$/C. The carrier concentration is about one order of magnitude higher than $Bi_4O_4S_3$ [7-8].

## 4. Conclusion

In conclusion, the newly discovered $LaO_{0.5}F_{0.5}BiS_2$ superconductor is synthesized in single phase and its bulk superconductivity is established below 2.7K from both magnetization and magneto-transport measurements. The normal state resistivity is semiconducting, implying a possibility for further increase in $T_c$ with optimized doping and chemical pressure. The *H-T* phase diagram shows linear variation of upper-critical field as a function of temperature near $T_c$ and $H_{c2}(0)$ is estimated to be ~ 19 kOe. The present communication is not only the timely reproduction of the ref.1 results but the sample quality is slightly improved in terms of phase purity.

**Acknowledgments**

Authors from NPL would like to thank their Director Prof. R. C. Budhani for his keen interest in the present work. This work is supported by DAE-SRC outstanding investigator award scheme on search for new superconductors. Anuj Kumar, Rajveer Jha, Shiva Kumar and Anand Pal are thankful to *CSIR*-India for providing the financial support during their research. Shruti and J. Saha acknowledge UGC for research fellowships. S. Patnaik thanks *AIRF*, JNU for the *PPMS* facility.

**Figure Caption**

**Figure 1 (a):** Observed and calculated (*solid lines*) XRD patterns of $LaOBiS_2$ and $LaO_{0.5}F_{0.5}BiS_2$ compounds at room temperature.

**Figure 1 (b):** Representative unit cell of $LaOBiS_2$. Color code Violet-Bi, Green-La, Yellow-S and Red-O/F

**Figure 2:** *DC* magnetization (*ZFC* and *FC*) plots for $LaO_{0.5}F_{0.5}BiS_2$, measured in the applied magnetic field, $H = 30$ Oe. Inset shows isothermal *MH* curve of the sample at 1.8K.

**Figure 3:** Resistivity vs. temperature ($\rho$-$T$) behavior of $LaO_{0.5}F_{0.5}BiS_2$ in various applied fields of 0, 0.2, 0.4, 0.6, 1.6, 2.2 *k*Oe in superconducting region; upper inset shows the zero field $\rho$-$T$ in extended temperature range of 2-300K for $LaOBiS_2$ and $LaO_{0.5}F_{0.5}BiS_2$; lower inset shows the temperature dependence of upper critical and irreversibility fields of $LaO_{0.5}F_{0.5}BiS_2$.

**Figure 4:** Hall resistivity is plotted as a function of magnetic field at 10K.

**Table1**

| *Atom* | *x* | *y* | *z* | *site* | *Occupancy* |
|---|---|---|---|---|---|
| $LaOBiS_2$ | | | | | |
| La | 0.25 | 0.25 | 0.090(1) | 2c | 1 |
| Bi | 0.25 | 0.25 | 0.632(2) | 2c | 1 |
| S1 | 0.25 | 0.25 | 0.371(6) | 2c | 1 |
| S2 | 0.25 | 0.25 | 0.819(2) | 2c | 1 |
| O | 0.75 | 0.25 | 0.00 | 2a | 1 |
| $LaO_{0.5}F_{0.5}BiS_2$ | | | | | |
| La | 0.25 | 0.25 | 0.103(2) | 2c | 1 |
| Bi | 0.25 | 0.25 | 0.622(1) | 2c | 1 |
| S1 | 0.25 | 0.25 | 0.362(1) | 2c | 1 |
| S2 | 0.25 | 0.25 | 0.825(4) | 2c | 1 |
| O | 0.75 | 0.25 | 0.00 | 2a | 0.5 |
| F | 0.75 | 0.25 | 0.00 | 2a | 0.5 |

**Table 1** Atomic coordinates, Wyckoff positions, and site occupancy for studied $LaO_{0.5}F_{0.5}BiS_2$

**Figure 1 (a)**

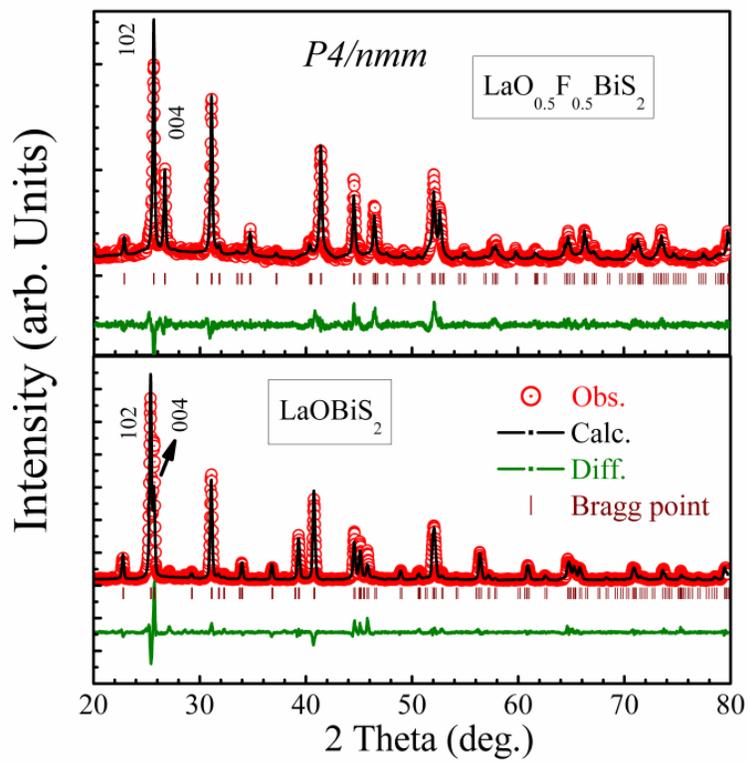

**Figure 1 (b)**

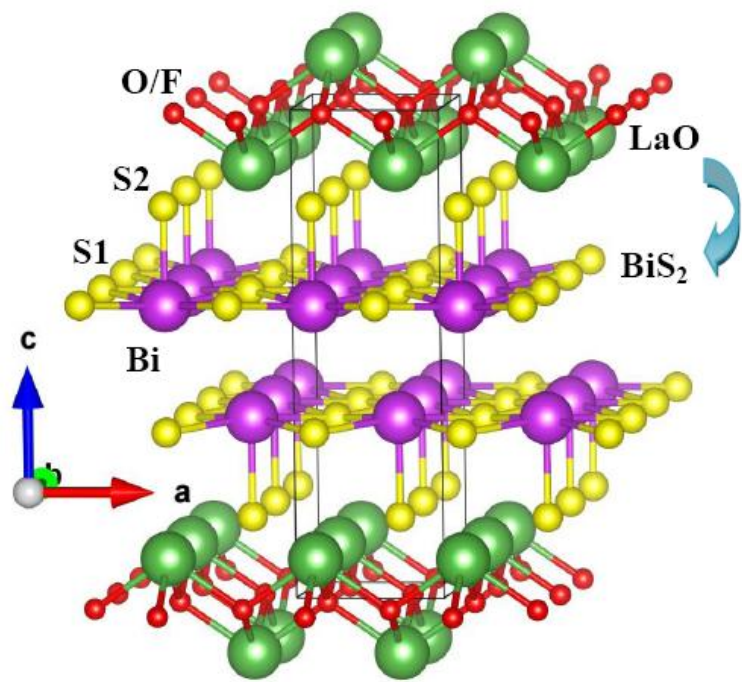

**Figure 2**

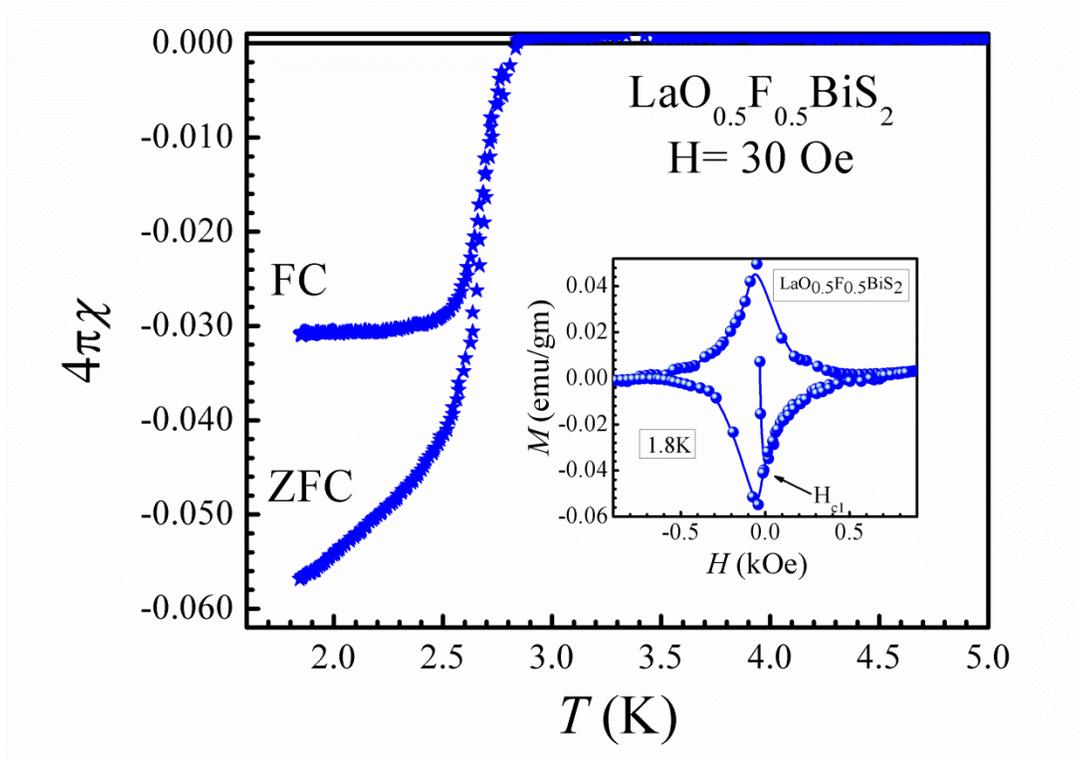

**Figure 3**

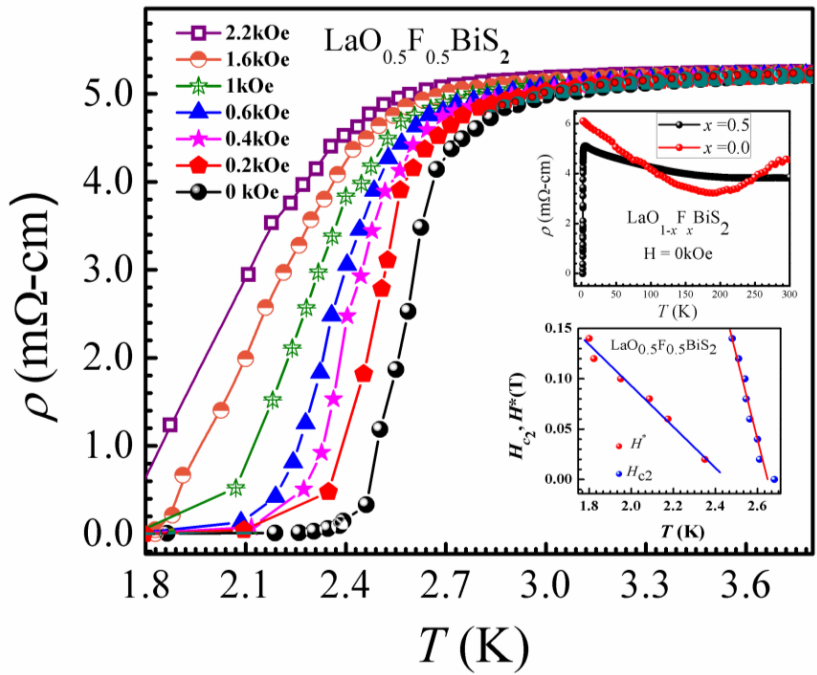

**Figure 4**

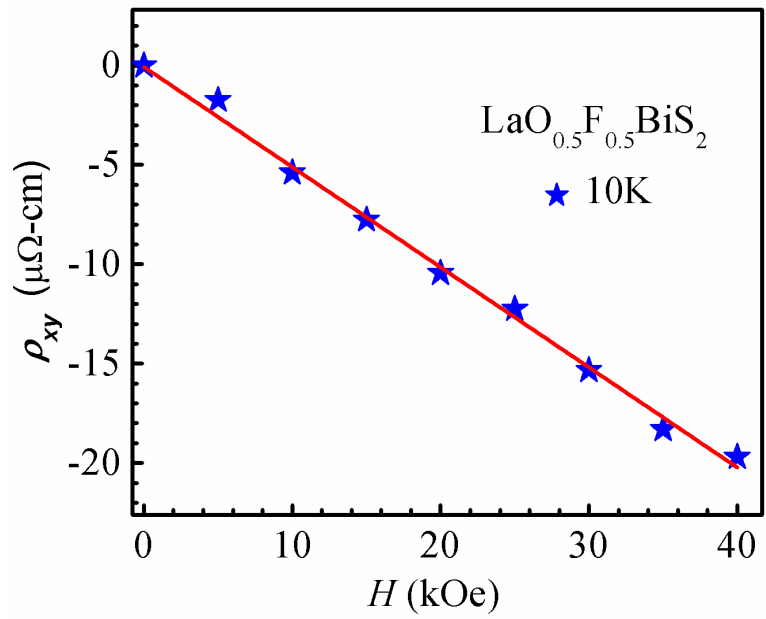